\documentclass[useAMS,usenatbib]{mn2e}
\usepackage{graphicx}

\usepackage{subfigure}

\title[Measurement of the Intracluster light at z $\sim$ 1]{Measurement of the intracluster light at $z\sim 1$}
\author[C. Burke et al.]{Claire Burke$^{1}$\thanks{E-mail: cb@astro.livjm.ac.uk (C.B.)}, Chris A. Collins$^{1}$, John P. Stott$^{2,1}$  and Matt Hilton$^{3}$\\
$^{1}$Astrophysics Research Institute, Liverpool John Moores University, Egerton Wharf, Brikenhead, Wirral, CH43 1LD, UK.\\
$^{2}$Extragalactic \& Cosmology Group, Department of Physics, Durham University, South Road, Durham, DH1 3LE, UK.\\
$^{3}$Centre for Astronomy \& Particle Theory, School of Physics \& Astronomy, University of Nottingham, Nottingham, NG7 2RD.}
\begin{document}

\date{Accepted 2012 June 20. Received 2012 May 11; in original form 2012 March 16}

\pagerange{\pageref{firstpage}--\pageref{lastpage}} \pubyear{2012}

\maketitle

\label{firstpage}

\begin{abstract}
A significant fraction of the total photospheric light in nearby galaxy clusters is thought to be contained within the diffuse intracluster light (ICL), which extends 100s of kpc from cluster cores. The study of the ICL can reveal details of the evolutionary histories and processes occurring within galaxy clusters, however since it has a very low surface brightness it is often difficult to detect. We present here the first measurements of the ICL as a fraction of total cluster light at $z\sim1$ using deep {\it J}-band (1.2 $\micron$) imaging from HAWK-I on the VLT. We investigate the ICL in 6 X-ray selected galaxy clusters at 0.8$\le z \le$1.2 and find that the ICL below isophotes $\mu_{J}$ = 22 mag/arcsec$^{2}$ constitutes 1--4\% of the total cluster light within a radius $R_{500}$. This is  broadly consistent with simulations of the ICL at a similar redshift and when compared to nearby observations  suggests that the fraction of the total cluster light that is in the ICL has increased by a factor 2 -- 4 since z$\sim$1. We also find the fraction of the total cluster light contained within the Brightest Cluster Galaxy (BCG)  to be 2.0--6.3\% at these redshifts, which in 5 out of 6 cases is larger than the fraction of the ICL component, in contrast to results from nearby clusters. This suggests that the evolution in cluster cores involves substantial stripping activity at late times, in addition to the early build up of the BCG stellar mass through merging. The presence of significant amounts of stellar light at large radii from these BCGs may help towards solving the recent disagreement between the semi-analytic model predictions of BCG mass growth (e.g. De Lucia \& Blaziot, 2007) and the observed large masses and scale sizes reported for BCGs at high redshift.

\end{abstract}


\section{Introduction}\label{intro}

Dominated by dark matter and with masses rising above $10^{15}\,$M$_{\odot}$, galaxy clusters are ideal regions for testing our understanding of astrophysical processes and revealing insights into the formation and evolution of structure in the Universe. The ubiquitous presence of hot gas and hundreds of galaxies means they can be studied out to high redshift in optical, X-ray and now SZ surveys, and used, not only as probes of the hierarchical evolution process, but also to measure cosmological parameters - for a recent review see Allen et al. (2011). 

In most clusters the brightest cluster galaxy (BCG) dominates the photospheric stellar light in the cluster core, with the BCG located close to the peak of the X-ray emission. For example, Stott et al. (2012) find that the average BCG offset from the cluster X-ray centroid is only 0.03R$_{500}$, a result consistent with Lin \& Mohr (2004), who find that for local massive clusters the BCG lies at an average distance of only  $\simeq$ 15 kpc from the cluster centroid - a figure which may rise to more like 50 kpc at $z\geq1$ (Fassbender et al., 2011).

Recent work on the mass and size evolution of BCGs has provided a challenge for current cosmological models and simulations. BCGs have been found to have significantly larger masses (Whiley et al., 2008; Collins et al., 2009; Stott et al., 2010) at high redshift than is predicted by semi-analytic models (e.g., De Lucia and Blaizot, 2007). These observations have demonstrated that BCGs have already undergone the majority of their mass evolution by $z$=1, contrary to the predictions of simulations which suggest a tripling in mass since this time. It has also recently been observed that the scale sizes of BCGs show very little increase since z$\geq$1 (see Stott et al., 2011), contrary to observations of passively evolving massive galaxies in the field which are observed to increase in  scale size by 2--5 times since z$\sim$2 without showing any significant mass increase. Whilst simulations of massive ellipticals show that this increase could be caused by minor mergers (e.g., Naab et al., 2009), there is no need for as many interactions in BCGs since z$\sim$1 to explain their modest size increase.

The notion of the intracluster light (ICL) first arose from observational results  that showed the surface brightness profiles of BCGs were extended, in excess of the predicted `classic' DeVaucouleurs (r$^{1/4}$) surface brightness radial profile for elliptical galaxies (Matthews et al., 1967; Sersic, 1968; Shombert, 1988). Estimates of the fraction of cluster light contained in the ICL in nearby clusters range from 10--50\%, with the upper end set by the observations in the core of Coma (Bernstein et al., 1995). With the growing realisation of the ICL's potential importance there have been concerted efforts on both the observational and theoretical front; new deep observations of ICL have been made in clusters now covering a range of redshifts out to $z$=0.8 (Gonzalez et al., 2005, 2007; Rudick et al., 2006; Krick et al., 2006, 2007; Guennou et al., 2011; Toledo et al., 2011). In addition, recent simulations of the ICL evolution with cluster evolution (e.g., Conroy et al., 2007; Murante et al., 2007; Puchwein et al., 2010;  Rudick et al., 2011) open up the potential of new constraints on the assembly history in rich clusters. 

The wide range of measured ICL fractions is at least in part due to the numerous methods used to treat the BCG extended halo and the ICL as separate components (e.g. Gonzalez et al., 2005; Lauer et al., 2007). Although some authors fit BCGs with a double surface brightness profile (e.g. Gonzalez et al., 2005; Ascaso et al., 2011) and often find an extended outer component many times larger than the inner component,  these fits are often degenerate making these composite analyses difficult to interpret without dynamical information (Dolag et al., 2010). Furthermore the different definitions, assumptions and fitting procedures make literature estimates of characteristic scale-sizes of BCGs difficult to compare. 

In our recent paper (Stott et al., 2011) we examine the surface brightness profiles of BCGs at $z\sim1$ using {\it HST} data and see tantalising hints of an extended surface brightness profile (beyond a DeVaucouleurs r$^{1/4}$ profile). This has motivated us to acquire deep {\it J}-band observations with HAWK-I on the VLT to examine the diffuse ICL component in six $z$=0.8--1.22 clusters. We present here the first search for the ICL in galaxy clusters at these redshifts and compare our results with the cosmological simulations of the ICL by  Ruddick et al. (2011) who use surface brightness thresholding to distinguish the ICL from other galaxy components, thereby circumventing some of the complexities resulting from the different parameterisations  of the extended light components discussed above.

The structure of this paper is as follows: in \S~\ref{data} we describe our data, observations and reduction; in \S~\ref{method} we describe the methods used in our measurements; in \S~\ref{results} we present our results and discuss possible sources of systematic error; and \S s~\ref{discussion} and~\ref{conclusions} contain the discussion of the results and the conclusions drawn respectively. All magnitudes described in this paper are in the Vega magnitude system unless otherwise stated. Throughout this paper we adopt a $\Lambda$CDM cosmology with $H_0$ = 70 km s$^{-1}$Mpc$^{-1}$, $\Omega_M$ = 0.3, $\Omega_{\Lambda}$ = 0.7.

\section{Data}\label{data}
Our sample consists of 6 of the most distant, spectroscopically confirmed, X-ray luminous galaxy clusters. The clusters are in the redshift range $0.8 \le z  \le 1.22$ and their relevant properties are listed in Table~\ref{clustertab}. The clusters were imaged using HAWK-I (High Acuity Wide field K-band Imager) sited at the Nasmyth focus of UT4 (Pirard et al., 2004).  The HAWK-I camera consists of four HAWAII 2RG $2048\times2048$ pix chips and the total field of view of HAWK-I is 7.5$\times$7.5 arcmin. Each detector covers 217 arcsec on each side and the mosaic has 15 arcsec gaps between the individual chips. The pixel scale of the detectors is 0.106 arcsec/pix. 

Observations took place in service mode between February and April 2011, the observing nights and corresponding seeing conditions are listed in Table \ref{obs_tab}. The observations were split into 1 hour blocks to allow for ease of scheduling.
During each 1 hour block the cluster was centered on each of the 4 chips in sequence and a 10 minute exposure was taken on each chip, to allow sufficient depth to be reached and to cover the gaps, resulting in a total of 40 minutes of exposure per block. Each of the 10 minute exposures was made up of 15 lots of $4\times 10$ sec exposures, jittered in a random pattern inside a box of size $60\times60$ arcsec (average offset $\sim$30 arcsec). The large $60\times60$ arcsec random jitter was chosen due to the large size of the BCG halo (radii typically $\sim$ 13 arcsec at z=1, equivalent to $\sim$ 100 kpc).
Observations were performed in the {\it J}-band (central wavelength 1.2 $\micron$) with a total on source integration time of between 2.1 to 4.2 hours. Dark and twilight flat field images were taken at the end of each night of observing. 

The minimum angular distance between the Moon and any cluster observed was set at 40 degrees separation and observations were performed during grey time.
The observations were scheduled in `Band B' sky conditions (second highest tier). The observations were taken at an air mass $<$ 2, and during the observations the seeing varied between 0.5 and 0.9 arcsec. Photometric observing conditions were not requested in our application and consequently only a third of the observations were taken in photometric conditions, with the rest of the data having only thin cirrus sky transparency. However we obtained at least one night of photometric data for each cluster to which all the non-photometric data was normalised, and a comparison of the zero points from photometric and non-photometric nights indicated zero point shifts of $<$ 0.15 mag in the measured magnitudes of the stars normalised to the  2MASS (Two Micron All Sky Survey) catalogue. 
 
\begin{figure*}
  \includegraphics[width=18cm]{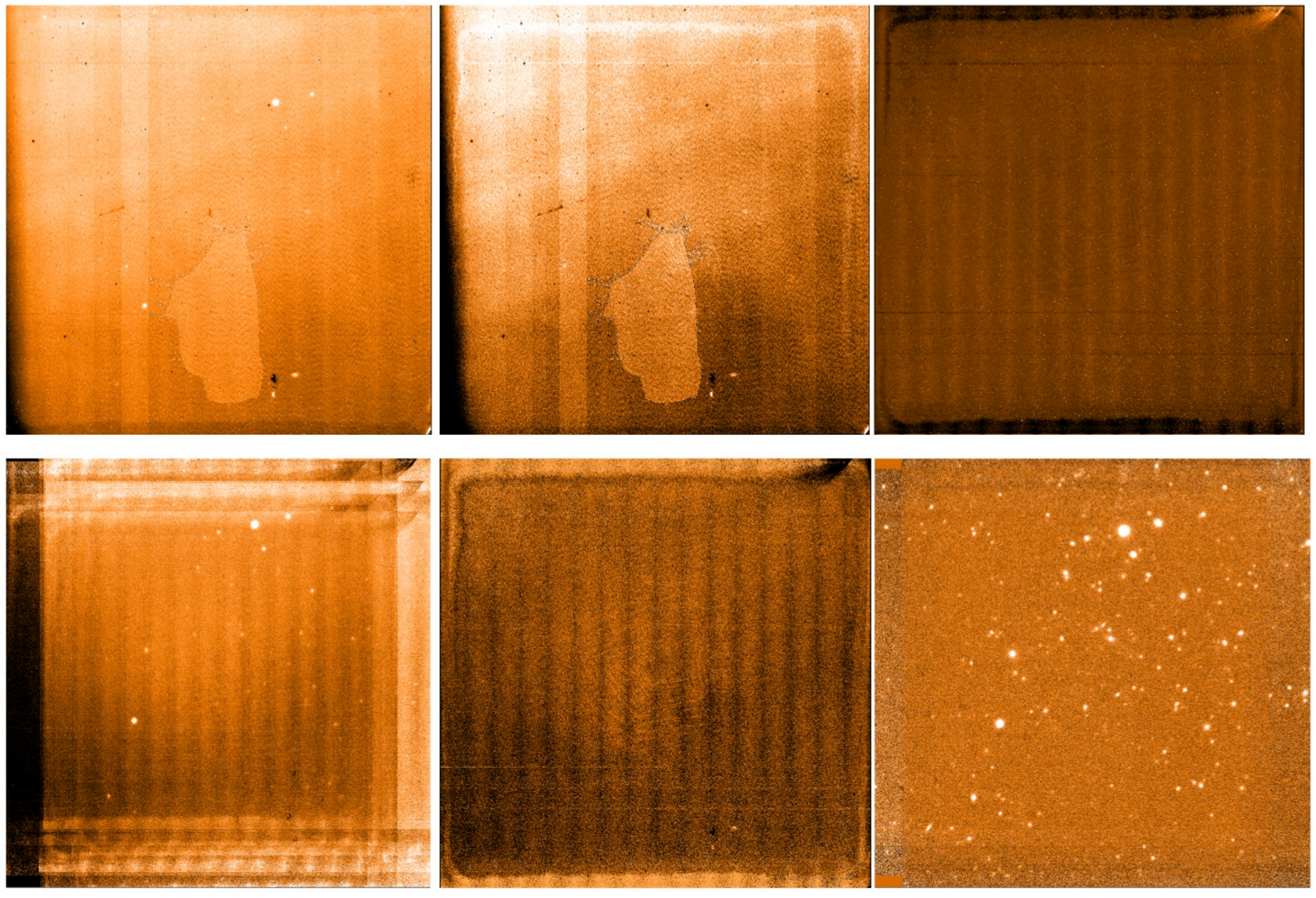}
  \caption{An example of the steps in the data reduction process. Top row: the raw data (top left), the twilight flat field (top centre), the dark frame (top right). Bottom row:  the data after basic dark frame removal and twilight flat fielding (bottom left), the resultant sky calculated from the running-median technique that is subtracted (bottom centre) and the data after all the reduction described above (bottom right). The cluster shown is  RDCS 1317 when observed for a quarter of an observing block on Chip 1. The size of each individual image is 217" x 217", equivalent to 1736 x 1736 kpc at z=1.}
  \label{raw_redn}
\end{figure*}

The data were reduced using the HAWK-I data reduction pipeline {\sc EsoRex}. The data underwent basic reduction by calibration with a dark frame, bad pixel mask and a twilight flat field frame. The basic reduced data were then sky subtracted using the sky subtraction package in {\sc EsoRex}, which computes a sky frame for each exposure from a running median of several jittered exposures. This individual computed sky frame is then subtracted from each individual exposure. The sky subtracted data were then median combined to create a fully reduced individual observing block of all 4 chips. An example of the raw, partially reduced and fully reduced data along with the twilight flat field, dark frame and subtracted sky frame are shown in Figure \ref{raw_redn}. 

\begin{figure*}
  \includegraphics[width=18cm]{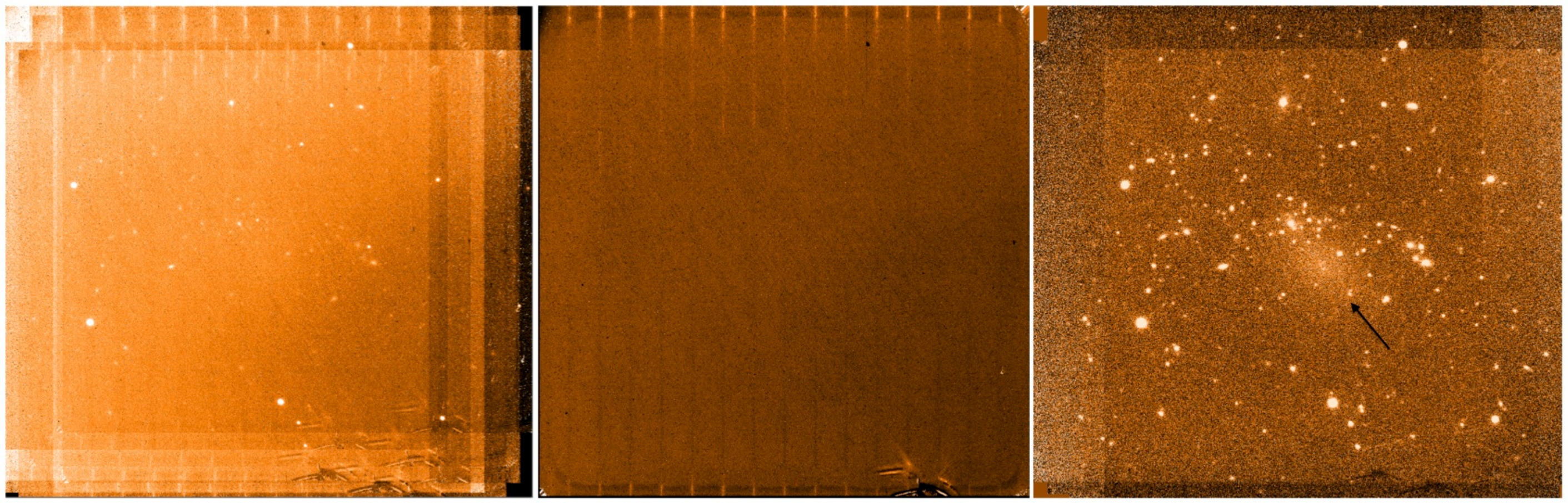}
  \caption{A model surface brightness profile is added into the data after it has undergone basic calibration by dark frame and twilight flat subtraction (left).  The sky that is subtracted (centre) shows no visible evidence of the profile (despite an rms pixel-to-pixel variation of 0.045 counts compared to the flux of the model ICL $\sim1.0$), while the model surface brightness profile is seen almost fully recovered in the sky subtracted data (right) at the position indicted. The cluster shown is MS 1054 when observed for a quarter of an observing block on Chip 2. The size of each individual image is 217" x 217".}
  \label{mod_sky_sub}
\end{figure*}

Low surface brightness components, like the ICL, can sometimes be unintentionally removed as sky during data reduction. In order to check that the ICL was not removed due to the finite jitter size during the sky subtraction we placed a model ICL profile into the data after dark current removal and twilight flat fielding but before the sky subtraction was performed. The model profile used was a Sersic profile with Sersic index n=1, corresponding to a flat, exponential disk-like profile (see Graham \& Driver (2005) for a detailed description of the Sersic profile), with a mean surface brightness $\sim$21 mag/arcsec$^2$ and a half light radius of 300 pixels ($\sim$250 kpc at z=1). The co-ordinate of the model profile was placed in each exposure so that it matched the jitter pattern offsets, effectively having the same World coordinates as the cluster centre. The sky subtraction was run over these images in the same way as for the normal data reduction. In several trials we found that a minimum of 83\% of the original flux in the model profile was recovered in the reduced data in an annulus between 100--300 pixels (85--250 kpc) and a minimum of 84\% was recovered in an annulus between 300 -- 500 pixels (250--425 kpc). For a visual comparison we show an example of the data with a model profile included in Figure \ref{mod_sky_sub}. As a sanity check we performed a similar analysis using a stellar profile placed near the centre of the clusters, recovering 95--96\% of the total stellar light. We conclude that our running-median sky subtraction technique recovers the bulk of a flat extended profile typical of an ICL covering the core of the clusters.
  
The reduction of the observations for each cluster from the observing blocks produced between 3 and 7 images for each cluster. These images were then median combined using the {\sc IRAF} function {\sc imcombine}. The surface brightness levels reached at $1\sigma$ for each cluster are shown in Table~\ref{clustertab}.

\begin{table*}
 \centering
  \begin{minipage} {15cm}
  \caption{Clusters observed with HAWK-I, all magnitudes are in the Vega system.}
  \label{clustertab}
  \begin{tabular}{@{}lcccccc@{}}
  \hline
   Name & RA & Dec &Redshift & Integration time & Limiting magnitude & Cluster mass\footnote{From Stott et al. (2010) (see references within)}\\
   &&&& Hours & J mag/arcsec$^{2}$ & x $10^{14} M_{\odot}$ \\
 \hline
\hline
   CL J0152 & 01 52 41 & -13 57 45 & 0.83 & 3.5 & 24.6  $\pm$ 0.2 & 4.5 $\pm$ 2.7 \\
   XLSS J0223 & 02 23 03 & -04 36 22 & 1.22 & 3.5 & 24.5  $\pm$ 0.1 & 1.8 $\pm$ 0.9 \\
   XLSS J0224 & 02 24 00 & -03 25 34 & 0.81 & 2.1  & 24.2  $\pm$ 0.1 & 2.3 $\pm$ 1.4 \\ 
   RCS J0439 & 04 39 38 & -29 04 55 & 0.95 & 3.5 & 24.5  $\pm$ 0.1 & 0.5 $\pm$ 0.4 \\
   MS 1054  & 10 57 00 & -03 37 27 & 0.80 & 4.2  & 24.3 $\pm$ 0.2 & 8.5 $\pm$ 4.9\\
   RDCS J1317 & 13 17 21 & ~29 11 18 & 0.81 & 2.8  & 24.2  $\pm$ 0.1 & 2.7 $\pm$ 2.9\\
 \hline
\end{tabular}
\end{minipage}
\end{table*}

\begin{table}
 \centering
  \caption{Observing nights and seeing conditions for our clusters observed with HAWK-I.}
  \label{obs_tab}
  \begin{tabular}{@{}lccc@{}}
  \hline
Cluster & Observing	& Date		& Seeing\\
	     &	 Block  	&			& (arcsec) \\
  \hline
  \hline
CL J0152	& 1	& 14/09/10	& 0.87\\
		& 2	& 14/09/10	& 0.57\\
		& 3	& 14/09/10	& 0.89\\
		& 4	& 18/09/10	& 0.56\\
		& 5	& 26/10/10	& 0.80\\
  \hline
XLSS J0223	& 1	& 16/10/10	& 0.58\\
		& 2	& 16/10/10	& 0.59\\
		& 3	& 16/10/10	& 0.56\\
		& 4	& 16/10/10	& 0.65\\
		& 5	& 11/12/10	& 0.88\\
  \hline
XLSS J0224	& 1	& 29/10/10	& 0.74\\
		& 2	& 29/10/10	& 0.69\\
		& 3	& 11/12/10	& 0.68\\
  \hline	
RCS J0439	& 1	& 14/09/10	& 0.70\\
		& 2	& 05/10/10	& 0.75\\
		& 3	& 16/10/10	& 0.57\\
		& 4	& 28/12/10	& 0.47\\
		& 5	& 22/02/11	& 0.73\\
  \hline		
MS 1054	& 1	& 06/01/11	& 0.46 \\
		& 2	& 22/02/11	& 0.81 \\
		& 3	& 22/02/11	& 0.57\\
		& 4	& 22/02/11	& 0.56 \\
		& 5	& 22/02/11	& 0.60 \\
		& 6	& 05/03/11	& 0.69\\
		& 7	& 05/03/11	& 0.62\\
  \hline		
RDCS J1317	& 1	& 22/02/11	& 0.82\\
		& 2	& 05/03/11	& 0.65\\
		& 3	& 12/03/11	& 0.51\\
		& 4	& 12/03/11	& 0.79\\
 \hline
\end{tabular}
\end{table}

\section[]{Measuring the Intracluster Light}\label{method}

Previous methods employed to estimate the ICL split broadly into two groups. Some adopt the approach of measuring both the  BCG and ICL together. This is often done because of the extended nature of BCGs, thus making it hard to distinguish between the two components. For example, as mentioned in the introduction, Gonzalez et al. (2005; 2007) fit the surface brightness profile of the BCG using two DeVaucouleurs profiles. A similar method was followed by Toledo et al. (2011), where all the galaxies in the cluster were masked out except for the BCG and the surface brightness profile of the extended BCG component was measured. This kind of measurement naturally  provides estimates of the fraction of cluster light which is contained in the BCG+ICL. 

The other group aims to determine the ICL contribution by removing light from galaxies and objects, including the BCG, above a surface brightness limit, and then integrating the remaining surface brightness across the whole cluster (Krick et al., 2006; 2007). Feldmeier et al. (2004) measure the flux in the ICL by masking out all the flux below a series of surface brightness levels and subtracting the remainder from the total flux in the cluster.

Our method for measuring the ICL is of the second type and was motivated so as to follow as closely as possible the predictions of Rudick et al. (2011) (hereafter R11), who use {\sc gadget-2} to simulate the evolution of 6 galaxy clusters with simulated cluster masses $\sim$ 10$^{14} M_{\odot}$, similar to the masses of our clusters (Table~\ref{clustertab}). With special emphasis on the dynamics of the ICL and galactic outskirts their method was first presented in Rudick et al. (2006), then overviewed and updated in R11. They predict the light from their simulated galaxy clusters below a series of surface brightness thresholds calculated between redshifts 0--2, covering the redshift range of our clusters. 
Hence we choose to measure the light in the galaxy clusters below a series of 7 surface brightness levels and count all the light below these levels as the ICL. Although this definition of the ICL does result in some loss of information regarding the contribution of the extremities of galaxies to the ICL, it is operationally effective requiring no separation of galaxies from ICL and makes comparison with the evolutionary predictions very straightforward.

Before any analysis was carried out all the stars and point sources in the cluster images were masked out. Point sources were identified as having a flux FWHM consistent with the average seeing for the combined image of that cluster ($0.5-0.9$ arcsec). We also masked out any objects with integrated magnitudes greater than that of the BCG to attempt to eliminate contamination from foreground galaxies (a more in depth discussion on the effects of contamination is found in \S~\ref{contamination}). We used {\sc SExtractor} (Bertin \& Arnouts, 1996) to measure the total isophotal flux in all the objects in the cluster (including the ICL) above the background, out to a cluster radius of $R_{500}$, corresponding to the radius at which the mean cluster density is 500 times the critical density. ICL `images' were made from the cluster data by placing an upper limit on the flux in the images corresponding to surface brightnesses in the {\it J}-band of: $\mu_J=18,19, 20, 20.5, 21, 21.5, 22$ mag/arcsec$^2$. This was done using the {\sc imarith} function in {\sc IRAF} - the total isophotal flux in these images above the background level was then measured using {\sc SExtractor}. 
The background used by {\sc SExtractor} was measured globally (BACKPHOTO type GLOBAL), so the sky was taken over the whole image and local variations were not subtracted. The spatial scale over which {\sc SExtractor} identified the background is set by the background mesh size parameter (BACK\_SIZE). In our analysis this was set to be larger than the average size of the objects in the images at 50 pixel$^2$ and was smoothed over 5 meshes.
The fraction of light contained in the ICL is thus defined as: ``the measured flux below a given surface brightness limit divided by the total flux in the cluster out to $R_{500}$''. 

The fraction of the total cluster light contained in the BCG was also measured. The surface brightness profiles for the BCGs were measured using the {\sc  ellipse} routine in the {\sc IRAF stsdas} package. To do this, first all other stellar and non-stellar objects in the cluster area are masked out to prevent contamination of the surface brightness profile. The flux is measured in ellipses at increasing radii from the centre of the BCG along its major axis. The 1D profile produced from this is then fitted with a model surface brightness profile using a chi-squared minimisation technique, similar to that described in Stott et al., 2011. The only significant difference between the method described in Stott et al. (2011) and the method used here is that the routine used here allows for the fitting of a point spread function (PSF) convolved profile, this allows us to fit the inner regions of the surface brightness profile - this correction was not needed for the {\it HST} data considered in Stott et al. (2011) due to the much smaller PSF size of {\it HST}. We fit the BCGs with DeVaucouleurs r$^{1/4}$ surface brightness profiles, which are shown in the Appendix.

To calculate flux contained in the BCG the best surface brightness profile fit is transformed into a 2D model of the BCG, with position angle and ellipticity matching that in the original data. To maintain consistency with measurement of the ICL, the flux in the model BCG is also measured using {\sc SExtractor} out to the same distance from the cluster centre ($R_{500}$) and the flux in the BCG is expressed as the fraction of the total flux in the cluster (see Table~\ref{results table}). 

To provide a conservative calculation of the errors the data were re-stacked so that a stacked image containing the cluster was made for each detector chip from all of the nights observed, i.e., all the observations of the cluster when it is positioned on chip 1 from each of 3, 4, 5 or 7 observing blocks (see Table \ref{obs_tab}) were stacked together. The error quoted was calculated from the variation between the measurement of the ICL on each of the 4 individual chips. Each individual chip stack reaches a shallower depth than the full stack of all the chips so we do not quote the ICL levels measured on these, merely the distribution of values between the measurements on each chip. The actual errors quoted are the equivalent 68\% percentile values (1 standard deviation for a Gaussian) from a Student's t-distribution of the ICL values measured on the 4 chips (Student's t-distribution being a more robust measure of the error for small number statistics, 4 in this case).

\begin{figure*}
  \includegraphics[width=18cm]{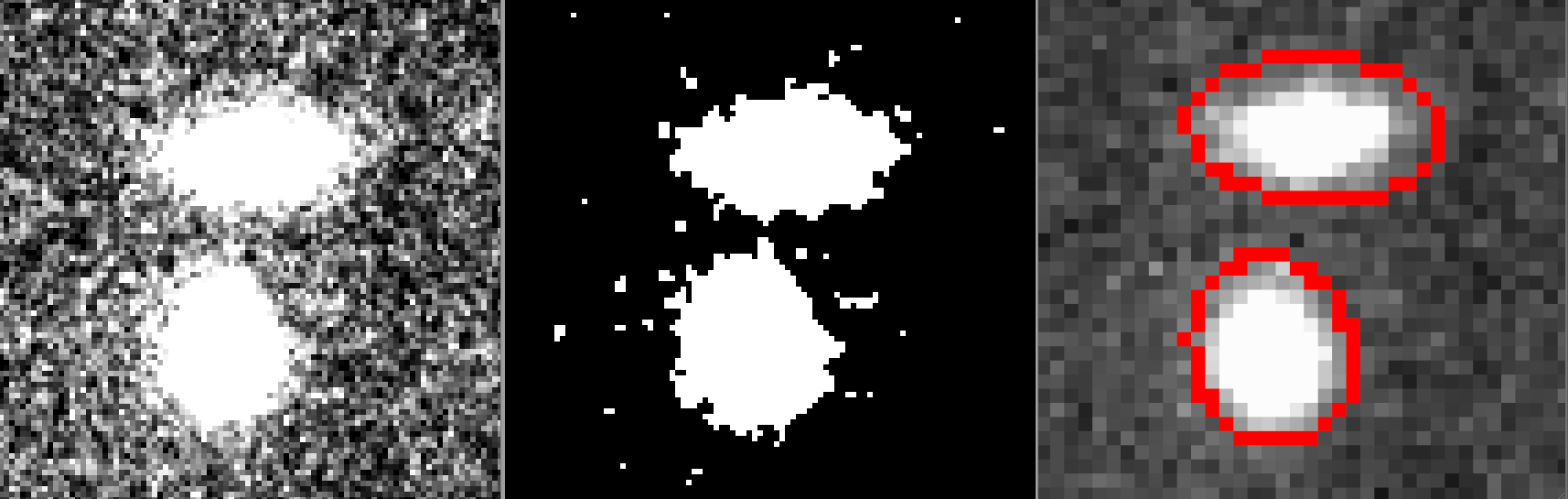}
  \caption{An illustration of the types of measurement of flux. Above are two elliptical galaxies in the cluster Cl J0152 (left). The surface brightness is measured above the flux level equivalent to $\mu_{J}$ = 20 mag/arcsec$^{2}$. Centre is a pixel map of all the pixels with flux above this level, in our measurement of the flux in the ICL this is the method we have used. Right is an illustration of the elliptical apertures generated by {\sc SExtractor} corresponding to this isophotal level. The size of these images is 6.4"x6.4" (50x50 kpc).}
  \label{iso_ell}
\end{figure*}

\section{Results}\label{results}

\begin{figure*}
\includegraphics[width =17cm]{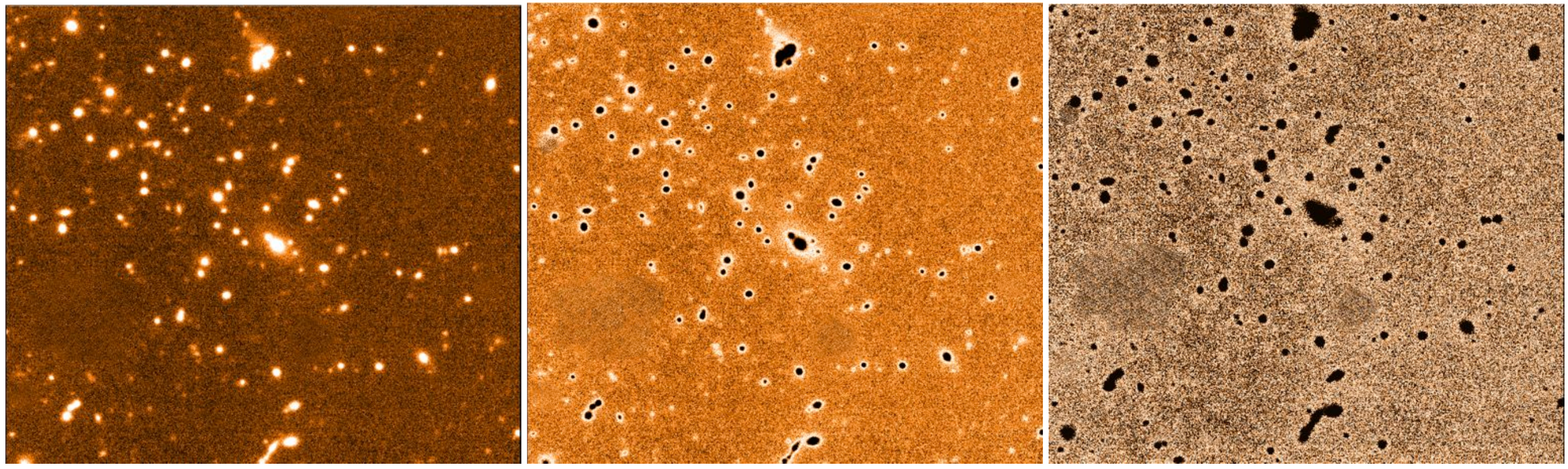}
\caption{The combined VLT image of CL J0152 (left). Proceeding images show the cluster with all light above given surface brightness levels blacked out to illustrate the light which is measured as ICL. The surface brightness limits correspond to $\mu_J$ = 19 (centre) and 21 (right) mag/arcsec$^{2}$. The size of the images shown here is 120"x110" (960 x 900 kpc).}
\label{0152_icl}
\end{figure*}

\begin{figure*}
\includegraphics[width =17cm]{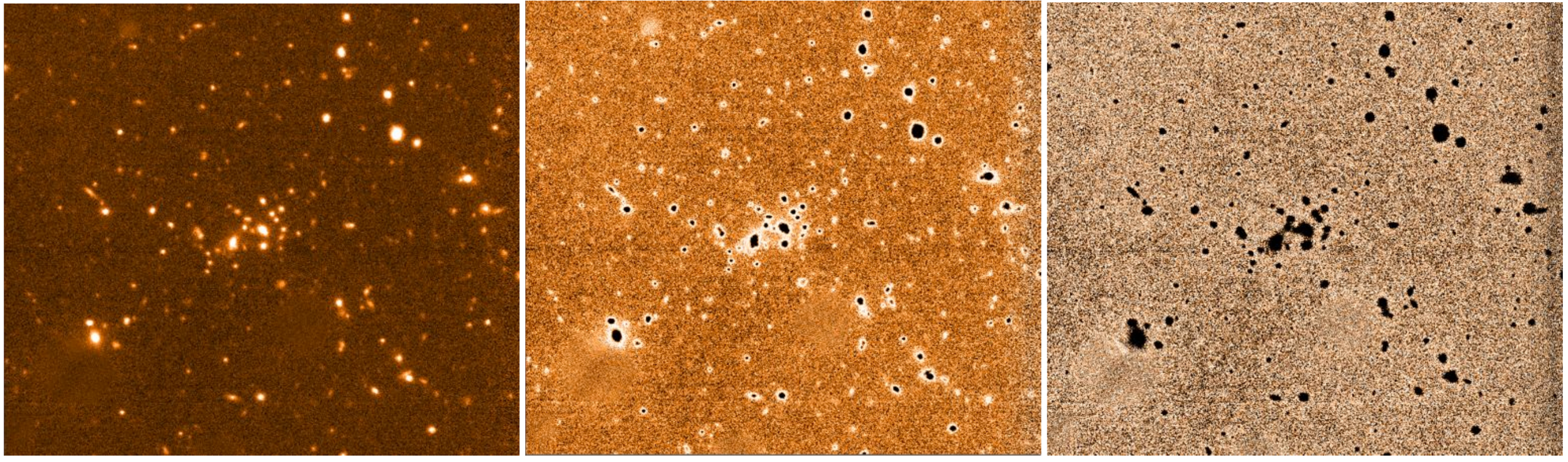}
\caption{As Figure~\ref{0152_icl} but for XLSS J0223. } 
\label{0223_icl}
\end{figure*}

\begin{figure*}
\includegraphics[width =17cm]{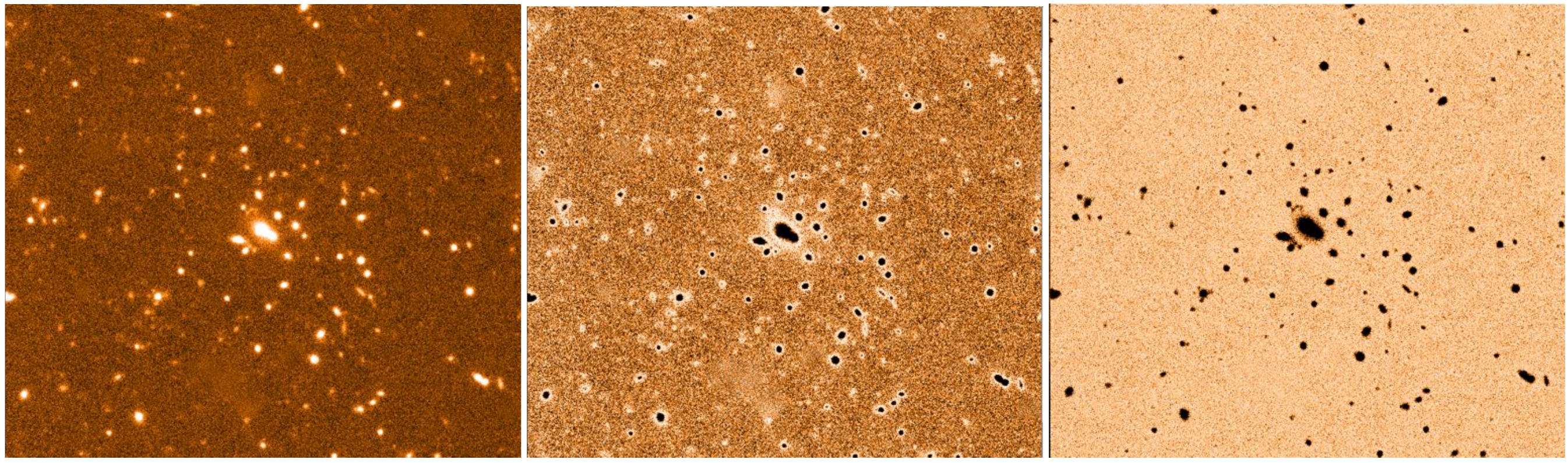}
\caption{As Figure~\ref{0152_icl} but for XLSS J0224.} 
\label{0224_icl}
\end{figure*}

\begin{figure*}
\includegraphics[width =17cm]{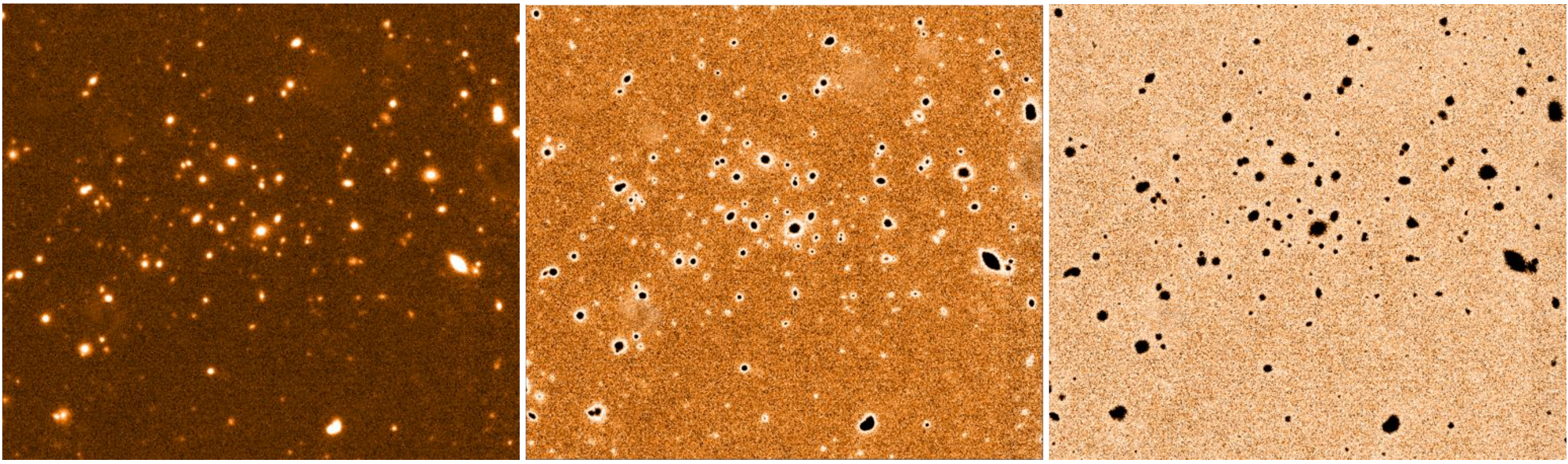}
\caption{As Figure~\ref{0152_icl} but for RCS J0439.} 
\label{0439_icl}
\end{figure*}

\begin{figure*}
\includegraphics[width =17cm]{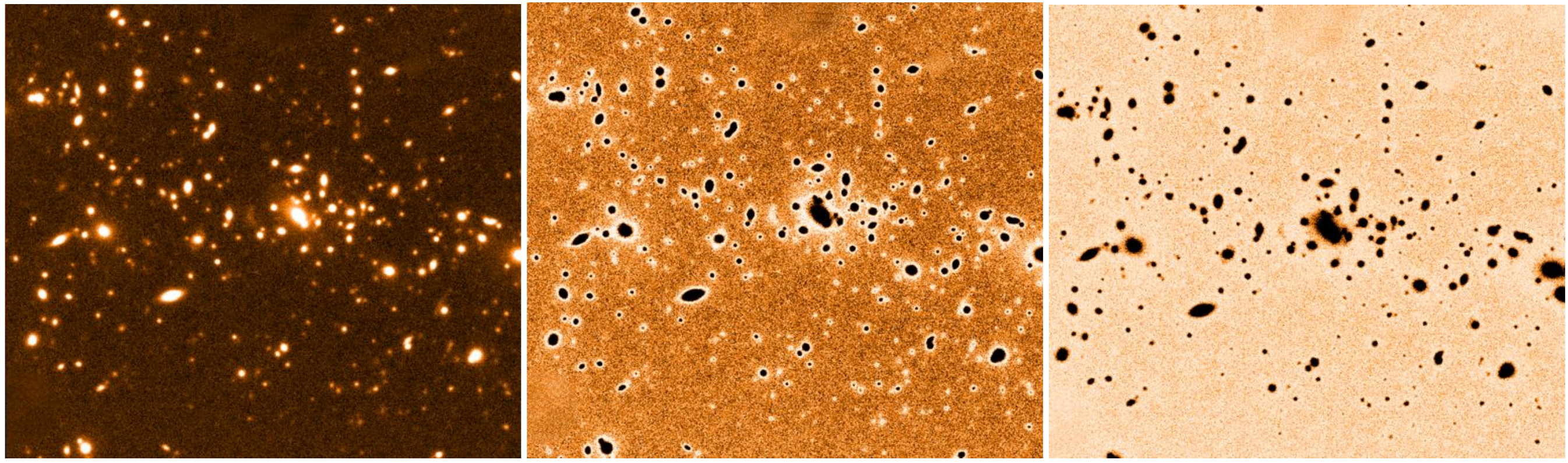}
\caption{As Figure~\ref{0152_icl} but for MS 1054.} 
\label{1054_icl}
\end{figure*}

\begin{figure*}
\includegraphics[width =17cm]{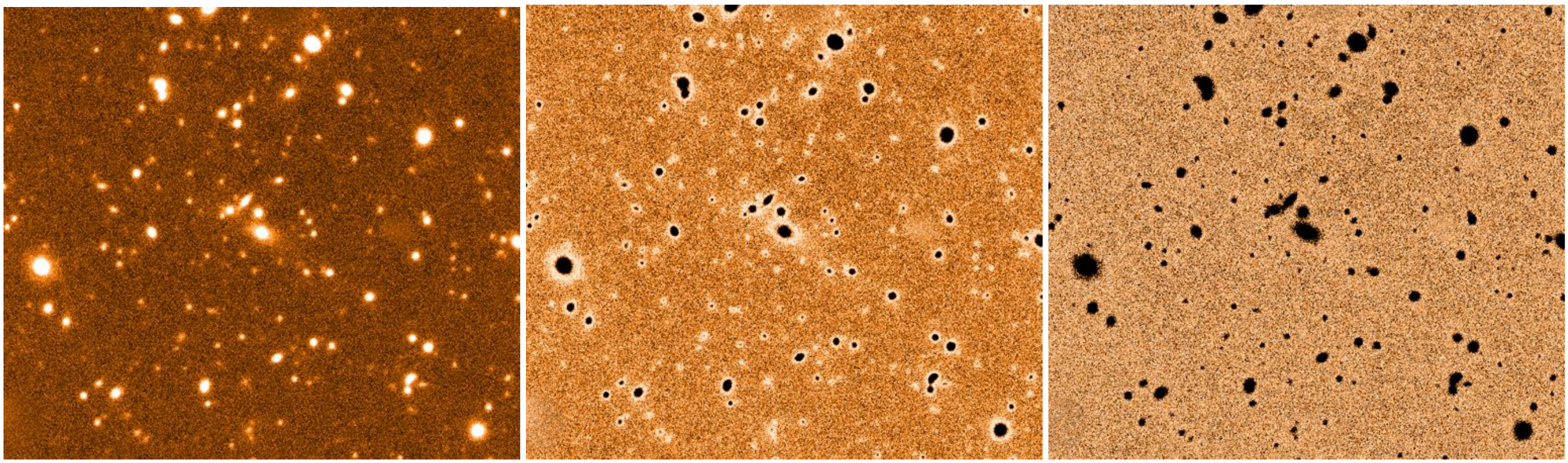}
\caption{As Figure~\ref{0152_icl} but for RDCS J1317. } 
\label{1317_icl}
\end{figure*}

Our results are summarised in Table~\ref{results table} where we list the percentage of light in the ICL at 7 different surface brightness limits in the {\it J} window. We make significant detections of the ICL in these high-$z$ clusters down to isophotal levels of $\mu_J$ = 22 mag/arcsec$^{2}$. At this limit we estimate  the ICL to contain $\sim$ 1 -- 4\% of the total light of the cluster. Illustrations of the measured ICL component at some of the surface brightness levels measured are presented in Figures~\ref{0152_icl} -~\ref{1317_icl} and clearly show that the detected extended  light follows the galaxy light very closely. We return to this point in \S~\ref{discussion}.

Also shown in Table~\ref{results table} is the percentage of the $J$-band light from the BCG estimated from the best DeVaucouleurs model fit, described above. For comparison with earlier work (e.g., Whiley et al., 2008; Aragon-Salamanca et al., 1998), we also measure the fraction of the $J$-band light in the BCG within a 50 kpc radius centred on the BCG in the reduced cluster images.

\begin{table*}
 \centering
  \caption{ICL percentages of the total cluster light in our cluster sample measured above surface brightness limits in the {\it J}-band. We also show percentages of the total cluster light contained in the BCG using two different measurements for comparison with previous studies.}
  \begin{tabular}{@{}lccccccccc@{}}
  \hline
     & \multicolumn{7}{c}{\% cluster light at surface brightness limit $\mu_{J}$} & \multicolumn{2}{c}{\% cluster light in BCG}\\
   Cluster & 18 & 19 & 20 & 20.5 & 21 & 21.5 & 22 & DeVaucouleurs & 50 kpc \\
    & & & &&&& & model& aperture\\

 \hline
\hline
   CL J0152 	& 84.7 $\pm$ 5.2 	& 73.6  $\pm$ 35.2 	& 51.6  $\pm$ 3.1	& 32.5  $\pm$ 2.9 	& 13.6  $\pm$ 1.9 & 5.4  $\pm$ 1.6 & 2.7  $\pm$ 0.4  & 3.8  $\pm$ 0.6  & 2.0 $\pm$ 0.2\\
   XLSS J0223	& 90.9 $\pm$ 4.2 	& 82.0 $\pm$ 21.9 	& 55.6 $\pm$ 2.5 	& 29.9 $\pm$ 1.0 	& 10.1 $\pm$ 0.8  & 3.6 $\pm$ 0.8  & 2.5 $\pm$ 0.4   & 2.0  $\pm$ 0.5  & 1.6 $\pm$ 0.2 \\
   XLSS J0224 	& 89.2 $\pm$ 27.4 	& 77.6 $\pm$ 12.4 	& 34.7 $\pm$ 11.6 	& 8.7 $\pm$ 2.0 	& 5.3 $\pm$ 0.9    & 2.6 $\pm$ 0.8  & 1.3 $\pm$ 0.4   & 6.3  $\pm$ 0.7  & 3.7 $\pm$ 0.3 \\ 
   RCS J0439 	& 80.0 $\pm$ 5.3 	& 70.1 $\pm$ 13.0 	& 47.2 $\pm$ 3.5 	& 25.7 $\pm$ 3.7 	& 9.3 $\pm$ 1.4    & 2.8 $\pm$ 2.6  & 1.5 $\pm$ 0.7   & 3.3 $\pm$  0.4  & 1.3 $\pm$ 0.2 \\
   MS 1054  	& 69.2 $\pm$ 7.9 	& 60.2 $\pm$ 15.1 	& 41.1 $\pm$ 4.2 	& 23.3 $\pm$ 1.5 	& 11.3 $\pm$ 2.7  & 6.2 $\pm$ 0.8  & 3.8 $\pm$ 0.2   & 4.7  $\pm$ 0.7  & 1.8 $\pm$ 0.3 \\
   RDCS J1317 & 78.5  $\pm$ 14.9 	& 66.0  $\pm$ 12.6 	& 32.4  $\pm$ 1.0 	& 12.4  $\pm$ 5.6 	& 4.6  $\pm$ 1.1   & 2.5  $\pm$ 0.4  & 1.5  $\pm$ 0.4 & 2.5  $\pm$ 0.5  & 1.3 $\pm$ 0.2 \\
 \hline
\end{tabular}
\label{results table}
\end{table*}

In order to compare our results with R11 we transform their rest frame $V$-band isophotes ($\mu_{V,rest}$) to predict our observed $J$ band ($\mu_{J, pred}$) using the expression; 

\begin{equation}
\mu_{J,pred}(z) = \mu_{V, obs}(z) +2.5log(1+z)^4 - K_{JV}(z).
\end{equation}

Here $2.5\,log(1+z)^4$ is the familiar surface brightness cosmological dimming term and $K_{JV}$ takes into account both the {\it k}-correction for accurate waveband comparison appropriate to the adopted Vega magnitude system (see Hogg et al., 2002) and the evolution of the stellar population. For the latter, R11 adopt a non-evolving fixed-age model of $\simeq10$ Gyr at $z=0.8, 1.0, 1.2$, which, as they admit, does not allow direct comparison with real clusters. However, to transform their predictions we also assume that the ICL is made up of an old stellar population, similar to the BCG's. This assumption of an old stellar population is justified based  on the findings of  other studies. Krick et al. (2006; 2007) observed the ICL in 10 clusters at $0.05 < z < 0.3$ and find the $B-r$ colour to be consistent with simple passive stellar evolution, lying on the same red sequence as the cluster elliptical galaxies and find the stellar populations in the ICL to have metallicities between Z$_\odot$ and 2.5Z$_\odot$ based on these colours. Further evidence that the two stellar populations have the same origin comes from the lack of any strong radial gradients in the $B-V$  colours of M87 in Virgo (Rudick et al. 2010). Finally, Zibetti et al. (2005) analyse the stack of 683 clusters between $z=0.2-0.3$ from SDSS-DR1 finding that the $g-, r- $ and $i-$band colours of the BCG and ICL are identical within the statistical uncertainties.

We transform the $\mu_V$ surface brightness values from the simulations of R11 to {\it J}-band surface brightness using the Bruzual \& Charlot (2003) {\sc galaxev} stellar population synthesis models. We used values for the formation redshift ($z_f$) of 3 and 5 and metallicity of the stellar population of Z$_\odot$ and 2.5Z$_\odot$. The range of $z_f$ between 3 and 5 is similar to the stellar ages found for high redshift BCGs (see Collins et al., 2009; Stott et al., 2010). The corrections for these two values of formation redshift and metallicity are listed in Table~\ref{corrections table}. Using the Bruzual \& Charlot models over this range of parameters gives rise to a mass-to-light ratio (M/L) $\sim$ 4 -- 6, which is consistent with the M/L of 5 used in the stellar evolution modelling of the ICL by R11 and characteristic of the older stellar populations expected in galaxy clusters.

\begin{table}
 \centering
  \caption{Total corrections applied for surface brightness dimming, $k$-correction and evolution correction for ICL at $z$=0.8, 1, 1.22, redshifts matching simulations we are comparing to, using stellar population models of differing redshift formation ($z_f$) and metallicity(Z).}
  \begin{tabular}{@{}lllccc@{}}
  \hline
  
 Redshift		&	 & 			& 0.8 	& 1	& 1.2\\
\hline
\hline				
2.5log(1+z)$^3$	& 			&	& 1.91	& 2.26	& 2.57\\
\\				
2.5log(1+z)$^4$	& 			&	& 2.55	& 3.01	& 3.42\\
\\				
Evolution   & z$_f$=3 &  Z$_\odot$			& 0.77	& 0.99	& 1.27\\
correction	 & z$_f$=3 & 2.5Z$_\odot$		& 0.80	& 1.07	& 1.39\\
		 & 	z$_f$=5  & Z$_\odot$		& 0.68	& 0.83	& 1.05\\
		 & 	z$_f$=5 & 2.5Z$_\odot$		& 0.72	& 0.85	& 1.14\\
\\				
K (colour)  &	z$_f$=3  & Z$_\odot$		& 2.0216	 & 1.9369 & 1.8527\\
correction	 & 	z$_f$=3 & 2.5Z$_\odot$		& 2.1375 	 & 1.9864 & 1.8660\\
		& 	z$_f$=5  & Z$_\odot$		& 2.0322	 & 1.9409 & 1.8575\\
		& 	z$_f$=5 & 2.5Z$_\odot$		& 2.1592 	 & 1.9994 & 1.8777\\
\\		
Total  	& 	z$_f$=3  & Z$_\odot$		& -0.8816 & -0.6669 & -0.5527\\
correction	 & 	z$_f$=3 & 2.5Z$_\odot$		& -1.0275 & -0.7964 & -0.6860\\
		& 	z$_f$=5  & Z$_\odot$		& -0.8022 & -0.5109 & -0.3375\\
		& 	z$_f$=5 & 2.5Z$_\odot$		& -0.9692 & -0.5894 & -0.4477\\
\hline

\end{tabular}
\label{corrections table}
\end{table}

In Figure \ref{rudick_plot} we plot the average of our results at our 5 faintest $\mu_{J}$ levels vs the ICL fraction along with the $\mu_{J, pred}$ predictions from the simulations of R11 at $z=0.8, 1.0, 1.2$. The simulations of R11 span a  range of $\mu_V$=25--27 mag/arcsec$^2$, corresponding to predicted near-infrared isophotes of $\mu_{J,pred}$ 24 -- 26.5 mag/arcsec$^2$, with only slight variations depending on the stellar population applied (see Table~\ref{corrections table}). The ICL fractions we find ($\sim$ 1 -- 4\%)  fall below the R11 predictions at 2--3 magnitudes brighter than the R11 surface brightness limit.

\subsection[]{Contamination}\label{contamination}

One of the potential problems in measuring ICL  fractions, particularly for clusters at such high redshift, is contamination from non-cluster galaxies. 
As mentioned in  \S~\ref{method}, we mask out any objects brighter than the BCG in an attempt to correct for contamination. Not correcting for this at all certainly affects our results; for example, if we measure the ICL fraction with just the point sources masked, leaving all the cluster and non-cluster galaxies in, the resulting fractions ICL at $\mu_J = 22$ mag/arcsec$^2$ reduce to about one half those given in Table~\ref{results table}. 

Attempts to correct for this contamination reported in the literature are somewhat patchy and dependent on other data available. Gonzalez et al. (2005) mask out point sources but do not discuss contamination from non-cluster members for their $z\leq0.13$ sample. Similarly, using various samples, Feldmeier et al. (2004), Zibetti et al. (2005) and Krick et al. (2006) conclude between them that contamination by non-cluster galaxies is not significant out to $z\simeq 0.3$. A more robust approach is taken by Toledo et al. (2011), who account for non-member contamination by calculating the luminosity function of the cluster from spectroscopically confirmed members then integrating the fitted luminosity function out to faint magnitudes to find the total luminosity of the cluster.

 In the same vein and in the absence of extensive redshift information for most of our sample, we estimate the effect of non-member galaxy contamination using the extensive spectroscopic survey of galaxies belonging to one of our clusters, CL J0152 (Demarco et al. 2010). These authors provide spectroscopic confirmation of cluster membership for 134 galaxies brighter than $K_S(AB) \simeq23$ (or $K_S=21.9$ in the Vega system), equivalent to $ J \simeq 23.7$ for an old stellar population at $z=0.8$. To test for the effect of contamination we masked out all the objects in our data within $R_{500}$ of the cluster centre which were not confirmed cluster members according to the Demarco et al. (2010) spectroscopic definition. Performing the ICL analysis again in an otherwise identical way, resulted in a $20\%$ increase in the ICL fraction (e.g. $3.2\pm 0.4$ \% c.f. $2.7\pm0.4$ \% at $\mu_J=22$ mag/arcsec$^2$). 
This is a strong test of the contamination of the total cluster light measured, rather than the total ICL. Although this tests affects the total cluster flux more strongly, we expect some of the flux in the ICL to be subtracted as well when non-cluster galaxies are removed.

Some increase is to be expected of course as bright cluster galaxies not targeted by Demarco et al. (2010) were also excluded in this test but in any case within the errors there is convincing evidence that contamination from non-cluster members is not a serious concern, as is illustrated by this example.

\subsection[]{Flat Fielding}\label{flat}

\begin{figure}
\includegraphics[width =9cm]{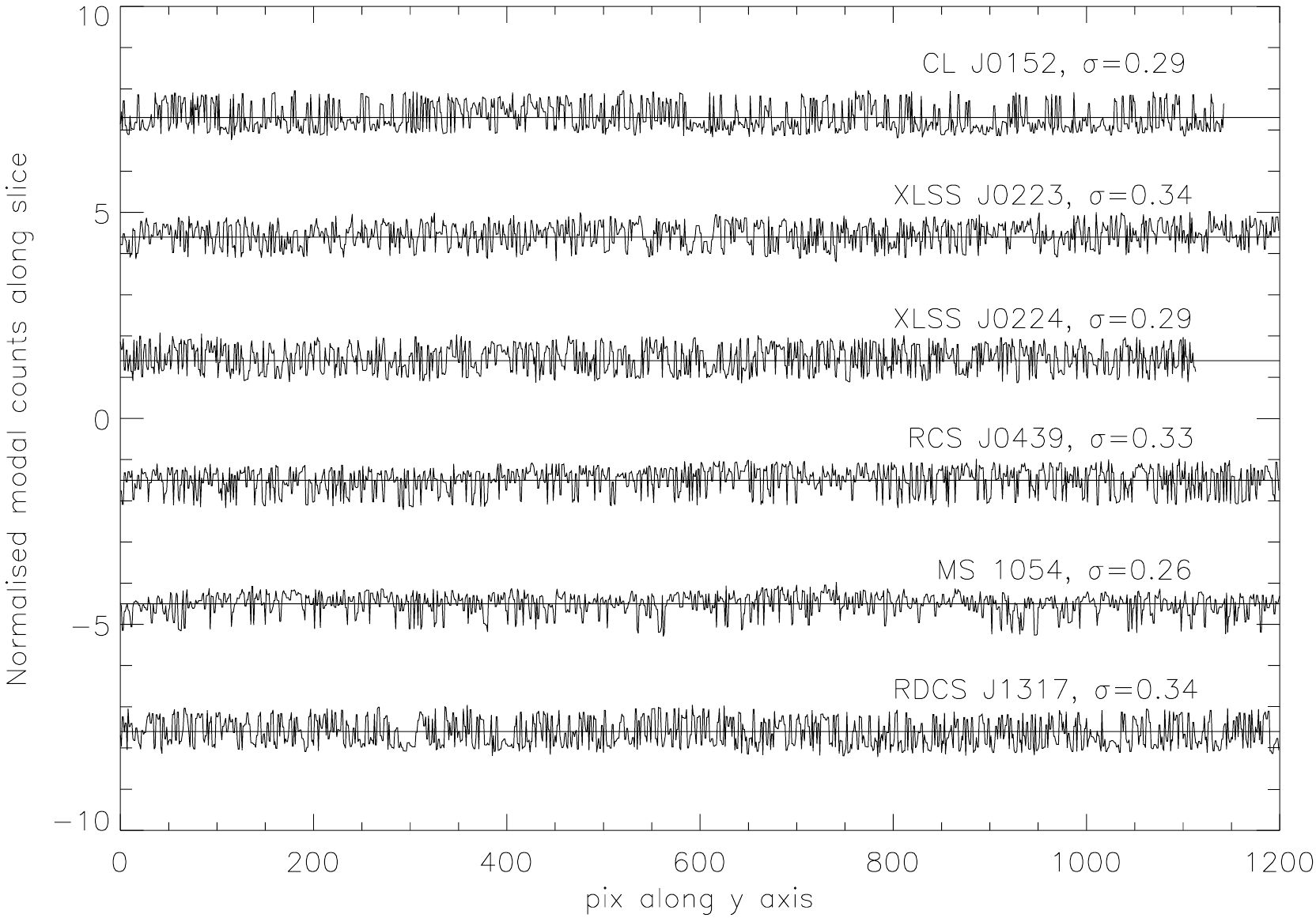}
\caption{The measured flat fielded background on our clusters along the y-axis of our cluster images. The mode was taken for each slice of image x-axis at the image y-values shown. The $\sigma$ values quoted are the standard deviation of the noise shown on these plots. The measurements have been offset along the y-axis of this plot for ease of viewing. A very similar plot is seen for slices along the x-axis of the images and is thus not included here.}
\label{flat_field}
\end{figure}

In order to recover accurate ICL values it is necessary to control the flat-fielding errors, particularly on large scales, as these are often the dominant source of uncertainty. The design of the 15 jittered exposures making up each observing block and described  in \S~\ref{data}, enables us to use the data reduction pipeline {\sc EsoRex} to compute the sky background and thus define a flat-field pattern for each cluster. In Figure~\ref{flat_field} we show the resulting sky background values (in normalised counts) on each of the final mosaiced images for our six clusters, calculated using the modal value of each slice. Fitting a simple straight line to these data, the largest linear gradient on the chip in either the x or y direction occurs in the mosaiced image of  XLSS J0223, which shows a change of 0.1 in the normalised counts over the central 1200 pixels, corresponding approximately to the region within $R_{500}$. Since the standard deviation in the background noise ranges from 0.26--0.34, we argue that the large-scale flat-field gradients in our sample are not significant and we do not attempt to correct for them or add a separate error component to account for them (for example, Feldmeier et al. 2004). In the same units as Figure~\ref{flat_field}, the ICL is clearly detected at $2-3 \sigma$ with the six measured ICL values ranging from 0.83--1.1, corresponding to surface brightnesses of $\mu_J=23.48 - 22.92$~mag/arcsec$^2$.

\begin{figure}
\begin{center}
\includegraphics[width =8.5cm]{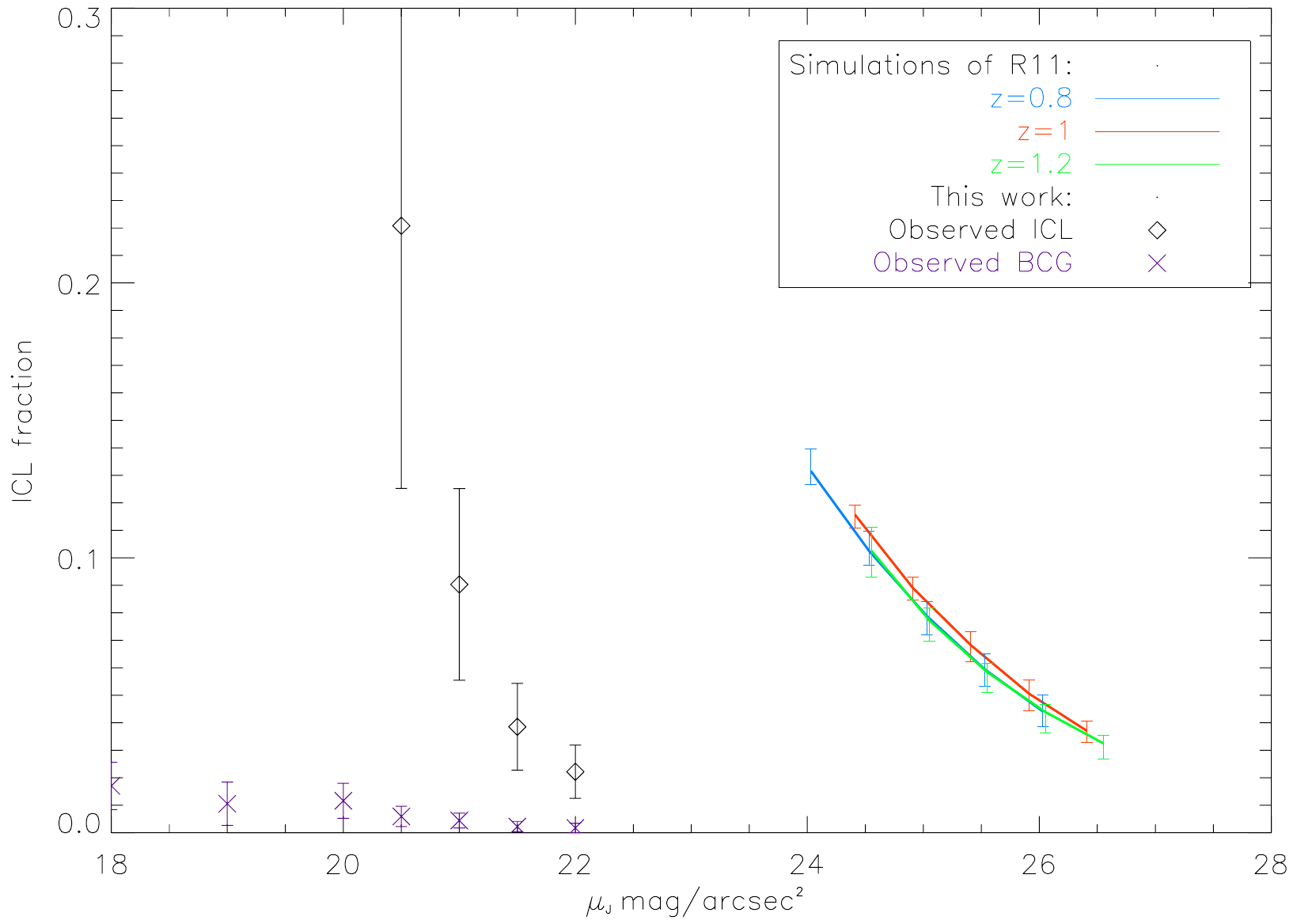}
\end{center}
\caption{ICL fraction vs surface brightness threshold in {\it J} at different redshifts. Coloured lines show the predictions of Rudick et al. (2011) at $z$=0.8 (green), $z$=1 (red), $z$=1.2 (blue). The average BCG fraction of the total cluster light at each surface brightness threshold is shown in dark blue (purple).}
\label{rudick_plot}
\end{figure}

\section{Discussion}\label{discussion}

At a surface brightness threshold limit of  $\mu_{J}$ = 22 mag/arcsec$^2$ we estimate that the ICL constitutes  1 -- 4\% of the total light of the clusters in our sample. These values are similar within the errors to those found in the simulations of R11, who find ICL fractions 4 -- 12\%, but are measured at a significantly brighter isophote, with R11 reaching  $\mu_{J}$ = 24--27 mag/arcsec$^2$ - an isophotal level at least 2 magnitudes fainter than our data. In Figure \ref{rudick_plot} we also see a steeper relation between the ICL fraction and surface brightness limit compared to R11. This may be evidence that there is less extended ICL than is predicted by the simulations at this redshift. 
A glance at Figures~\ref{0152_icl} --~\ref{1317_icl} indicates that the ICL is concentrated close to the cluster galaxies in the core rather than smoothly distributed. It may be that an extended component is below our sensitivity threshold, although it is known that measured ICL components closely follow the galaxy distribution in a number of low redshift clusters (Feldmeier et al. 2004).

As described in \S~\ref{intro}, due to the different definitions employed and the range of methods used to measure the ICL direct comparisons with other observations are difficult. Also there are a wide range of ICL fractions reported between different authors and within individual samples, nonetheless we point out some general trends here for work that uses a similar analysis to our own. For example, Feldmeier et al. (2004) measure the ICL in 4 non-cD clusters in the redshift range $z$=0.15 -- 0.2, using a similar surface brightness threshold technique, finding that the ICL below $\mu_{V}$ = 26 mag/arcsec$^{2}$ contains $16-28\%$ of the total cluster light, falling to $3-9\%$  below $\mu_{V}$ = 27 mag/arcsec$^{2}$. Krick et al. (2006, 2007) measure the ICL in 10 clusters at $z$ = 0.05--0.3 and find that the ICL below $\mu_{V}$ = 26 mag/arcsec$^{2}$ contains 4--21\% of the total cluster light. Zibetti et al. (2005) use the SDSS to measure the ICL in 683 clusters at $z$=0.2--0.3 by plotting surface brightness profiles of clusters out to 700 kpc, defining the ICL to be all the light below $\mu_{r}$= 25 mag/arcsec$^2$ and masking out light above this level.  Within 500 kpc they find that the ICL contains $10.9\pm5.0\%$ of the total cluster light, with the ICL being more centrally concentrated than the galaxies and the ICL fraction decreasing slowly with radius.
The ICL fractions we measure in our high-z clusters are generally lower than those found by other studies at low-z where similar ICL measurement methods have been used, which suggests a growth in the fraction of ICL of perhaps a factor of 2 -- 4 since $z$=1.

Insight into the mass assembly process at the centres of clusters comes from comparing the relative contributions from the BCG and ICL to the total light. From Table \ref{results table} the fraction of cluster light contributed by both the ICL and the BCG, measured as described in \S\ref{method}, ranges from 4.0 -- 8.5\%. This is smaller than the fractions reported by previous studies at low redshifts (e.g., Gonzalez et al., 2005, 2007; Zibetti et al., 2005; Toledo et al., 2011) and the predictions (Conroy et al., 2007), which find that the BCG+ICL constitutes as much as $\sim$ 33 -- 89\% of the total cluster light. However, whereas the majority of the BCG+ICL (up to 80\%) is in the ICL at low redshift, at z=1 we are seeing the biggest contribution to the BCG+ICL component coming from the BCG, constituting $\simeq$ 60\% of the BCG+ICL compared to a contribution of 40\% from the ICL. These results can be understood as a direct consequence of the stripping of surrounding galaxies in the cluster cores thus increasing the ICL component over time and further indicates that activity between galaxies in the centres of rich clusters may involve more stripping than merging, with matter from interactions ending up in the ICL or cD halo rather than centrally in the BCG. 

If the ICL does indeed show significant evolution over the same timescale where the BCG does not, it would be further evidence that these two components have separate evolutionary paths and separate origins. If the two distinct merging and stripping histories are correct, then one might well expect fossil groups, as the most dynamically evolved clusters, to exhibit large amounts of ICL surrounding their central high mass BCGs (see Harrison et al., 2012).

The quantity of extended halo light surrounding the central BCGs in clusters may go some way towards explaining why previous estimates of the stellar mass in BCGs at high redshift using {\it K}-band light as a mass proxy are significantly larger than predicted from semi-analytic models (Aragon-Salamanca et al., 1998; Whiley et al., 2008; Collins et al., 2009; Stott et al., 2010; 2011). This idea originates from a comment made in Whiley et al. (2008), that despite large BCG stellar masses at z$\sim1$, much stellar mass growth may be taking place outside of their fixed aperture magnitude measurements, made with a metric circular aperture of 37\,kpc diameter; similarly the aperture sizes in the Collins et al. (2009) and Stott et al. (2010) papers range from 11- 30\,kpc diameter. Testing this result we find (Table~\ref{results table}) that the contribution from the stellar light within a 50 kpc diameter aperture centred on the BCG is similar on average to the ICL contribution at $\mu_{J}$=22 mag/arcsec$^2$ measured outside that region, and constitutes about 50\% of the total estimated BCG light based on the 2D DeVaucouleurs model profile extrapolated to R$_{500}$. These simple comparisons indicate that indeed significant mass growth is likely to be occurring in the ICL compared to the BCG. Unfortunately the semi-analytic models of De Lucia \& Blaizot (2007), based on the Millennium Simulation and used for comparison in the BCG work above, do not tag the stripped stellar content in the ICL. More recent work by Puchwein et al. (2010) emphasises just this point. They have carried out high-resolution hydrodynamical simulations of clusters, including all known physical and feedback processes, to predict the relative build up of the ICL and central cluster galaxies. They find persistently high stellar fractions $\sim40\%$ in the ICL at late times, with the vast majority of stars in the main cluster haloes belonging to the ICL rather than the BCG -  a result which underlines the necessity of including both the BCG and ICL components when considering the stellar mass budget at the centres of clusters.  

The results of Gonzalez et al. (2007) for a sample of 24 clusters and groups at $z\leq0.13$ show an inverse correlation between the stellar ICL+BCG fraction and the cluster velocity dispersion, indicative of the somewhat counter intuitive result that the ICL component grows less efficiently 
in more massive host environments. A non-parametric Spearman rank correlation test applied to our data at the surface brightness limit of $\mu_{J}$ = 21 mag/arcsec$^{2}$ gives a coefficient of 0.37, with a significance of deviation from zero of 0.47, and at $\mu_{J}$ = 20 mag/arcsec$^{2}$ we find a Spearman rank coefficient of $-$0.26, with a significance of deviation from zero of 0.62. A null result is probably not surprising given our relatively small ICL+BCG fractions, although Zibetti et al. (2005) also conclude that the ICL fraction remains fairly constant (to within $\pm5\%$) between different cluster richnesses for their SDSS based sample of 683 clusters between $z$=0.2 -- 0.3. This possibly indicates selection differences between the samples, however the matter has yet to be settled.  

In our analysis we do not aim to distinguish between the ICL and the haloes of the cluster galaxies, but adopt a pragmatic ICL definition based simply on an isophotal threshold. However, as already mentioned, it is clear in Figures~\ref{0152_icl} -~\ref{1317_icl} that the extended component we detect closely follows the cluster galaxies, especially at higher surface brightness levels (e.g., $\mu_{J}$ =19 mag/arcsec$^2$ in Figures~\ref{0152_icl} -~\ref{1317_icl}), rather than constituting a diffuse component spread uniformly between the galaxies. Our data is not deep enough to detect the presence of a faint diffuse (uniformly spaced) ICL component, indeed our data would not detect the redshifted diffuse ICL component in Virgo at $\mu_{V} \geq 27$ mag/arcsec$^{2}$ for example (Rudick et al., 2010). As colour gradients in the outer parts of BCGs are shallow, dynamical studies using integral field spectroscopy on 8 metre or larger telescopes may be the most profitable way forward to separate the BCG halo and ICL components (Dolag et al., 2010).

\section{Conclusions}\label{conclusions}

We have detected and measured the fraction of cluster light that is in the ICL for 6 galaxies at $z\sim$1 using a simple definition of surface brightness threshold. We find that an extended component is detectable down to surface brightness levels of  $\mu_{J}\sim$22 mag/arcsec$^2$ measured within a radius R$_{500}$. At this level, the fraction of total measured cluster light in the ICL for our 6 clusters ranges between 1\% and 4\%, which is smaller than observations at lower redshift and similar or slightly below the predicted values at fainter isophotes, based on a similar ICL definition from the simulations of R11. This indicates that the ICL may have grown by a factor of 2 -- 4  since $z\sim$1, a scenario which is consistent with the idea of material being stripped from galaxies through mergers and close galaxy encounters.

In the context of the cosmological mass assembly problem of BCGs reported in the literature, the quantity of extended light is comparable to the centrally concentrated light from the BCGs. Taking into account both components is likely to ease the current discrepancy between the observed and predicted timescales of BCG assembly.

\section*{Acknowledgments}

We thank the anonymous referee for an excellent report which helped us improve the presentation of the paper and demonstrate the robustness of our results.
This paper is based on observations at the Very Large Telescope (VLT) of the European Observatory (ESO) in Chile.
We thank Craig Rudick for sharing his simulation data for our comparison and acknowledge Ivan Baldry for his helpful advice on the {\it k}-correction.
CAC acknowledges STFC for financial support from grant ST/H/002391/1. 

IRAF is distributed by the National Optical Astronomy Observatories, which are operated by the Association of Universities Research in Astronomy, Inc., under cooperative agreement with the National Science Foundation.

This publication makes use of data products from the Two Micron All Sky Survey, which is a joint project of the University of Massachusetts and the Infrared Processing and Analysis Center/California Institute of Technology, funded by the National Aeronautics and Space Administration and the National Science Foundation.

\appendix

\section[]{Surface brightness profiles of individual BCGs}

\begin{figure*}
  \includegraphics[width=17cm]{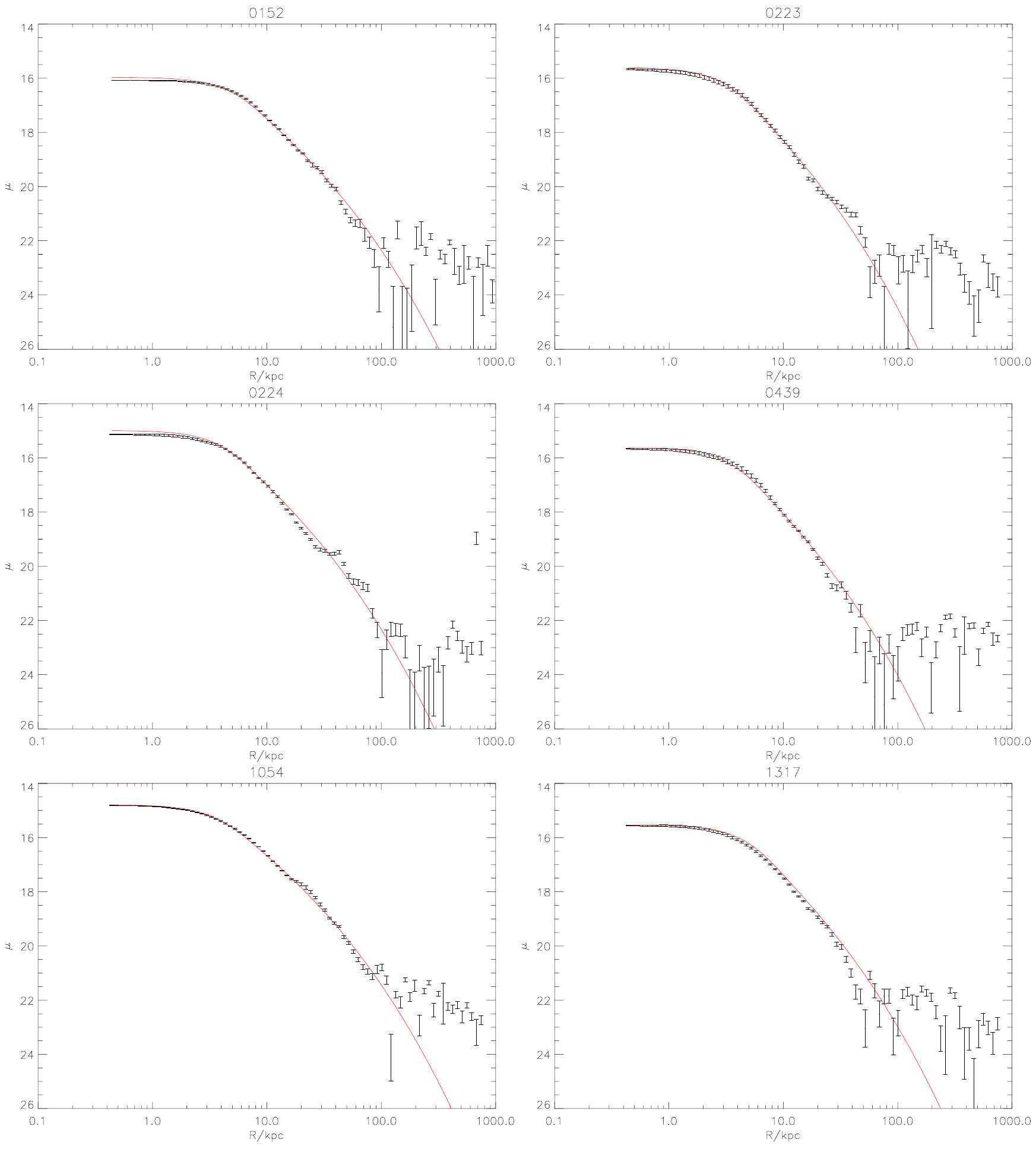}
  \caption{The 1D surface brightness profiles of the BCGs in our sample with best fitting Devaucouleurs profiles. The best fitting profiles are convolved with the observed PSF for each fit.}
  \label{individual SBs}
\end{figure*}

\label{lastpage}

\end{document}